\documentclass[fleqn,usenatbib]{mnras}

\usepackage[T1]{fontenc}
\usepackage{ae,aecompl}

\usepackage{graphicx}	
\usepackage{amsmath}	
\usepackage{amssymb}	
\usepackage{color}
\usepackage{algorithm}
\usepackage{booktabs}
\usepackage[noend]{algpseudocode}
\usepackage{float}
\usepackage{multibib}
\newcites{Appendix}{Appendix References}
\usepackage{natbib}
\usepackage[flushleft]{threeparttable}
\usepackage{mathrsfs}

\graphicspath{{Figures/}}

\DeclareMathOperator\erfc{erfc}
\DeclareMathOperator\Dmin{\mathcal{D}_\text{min}}
\DeclareMathOperator\Dmax{\mathcal{D}_\text{max}}
\DeclareMathOperator\D{\mathcal{D}}
\DeclareMathOperator\M{\mathcal{M}}
\DeclareMathOperator\Ml{\mathscr{M}}

\DeclareMathOperator\SF{\text{SF}}

\title[From 2D projections to 3D turbulent Mach number]{The relation between the turbulent Mach number and observed fractal dimensions of turbulent clouds}  

\author[J. R. Beattie et al.]{
James R. Beattie$^{1}$\thanks{E-mail: beattijr@mso.anu.edu.au}, Christoph Federrath$^{1}$\thanks{E-mail: christoph.federrath@anu.edu.au}, Ralf S. Klessen$^{2,3}$ and Nicola Schneider$^{4}$
\\
$^{1}$Research School of Astronomy and Astrophysics, Australian National University, Canberra, ACT 2611, Australia\\
$^{2}$Universit\"at Heidelberg, Zentrum f\"ur Astronomie, Institut f\"ur Theoretische Astrophysik, Albert-Ueberle-Str. 2, 69120 Heidelberg, Germany\\ 
$^{3}$Universit\"at Heidelberg, Interdisziplin\"ares Zentrum f\"ur Wissenschaftliches Rechnen, Im Neuenheimer Feld 205, 69120 Heidelberg, Germany \\
$^{4}$I. Physik. Institut, University of Cologne, D-50937 Cologne, Germany
}

\date{Accepted XXX. Received YYY; in original form ZZZ}

\pubyear{2019}

\begin{document}
\label{firstpage}
\pagerange{\pageref{firstpage}--\pageref{lastpage}}
\maketitle

\begin{abstract}
Supersonic turbulence is a key player in controlling the structure and star formation potential of molecular clouds (MCs). The three-dimensional (3D) turbulent Mach number, $\M$, allows us to predict the rate of star formation. However, determining Mach numbers in observations is challenging because it requires accurate measurements of the velocity dispersion. Moreover, observations are limited to two-dimensional (2D) projections of the MCs and velocity information can usually only be obtained for the line-of-sight component. Here we present a new method that allows us to estimate $\M$ from the 2D column density, $\Sigma$, by analysing the fractal dimension, $\D$. We do this by computing $\D$ for six simulations, ranging between $1$ and $100$ in $\M$. From this data we are able to construct an empirical relation, $\log\M(\D) = \xi_1(\erfc^{-1} [(\mathcal{D}-\Dmin)/\Omega] + \xi_2),$ where $\erfc^{-1}$ is the inverse complimentary error function, $\Dmin = 1.55 \pm 0.13$ is the minimum fractal dimension of $\Sigma$, $\Omega = 0.22 \pm 0.07$, $\xi_1 = 0.9 \pm 0.1$ and $\xi_2 = 0.2 \pm 0.2$. We test the accuracy of this new relation on column density maps from $Herschel$ observations of two quiescent subregions in the Polaris Flare MC, `saxophone' and `quiet'. We measure $\M \sim 10$ and $\M \sim 2$ for the subregions, respectively, which is similar to previous estimates based on measuring the velocity dispersion from molecular line data. These results show that this new empirical relation can provide useful estimates of the cloud kinematics, solely based upon the geometry from the column density of the cloud. 
\end{abstract}


\section{Introduction}\label{introduction}
The dynamical evolution of molecular clouds (MCs) in the interstellar medium (ISM) is determined by supersonic, compressible turbulent flows \citep{Larson1981,Solomon1987,Klessen2000,Heitsch2001,Ossenkopf2002,Elmegreen2004,Heyer2004,MacLow2004,Scalo2004,Krumholz2005,Ballesteros2007,McKee2007,RomanDuval2011,Padoan2014,Federrath2015}. The turbulent dynamics of the clouds plays a diverse and vital role in the star formation process by providing support against collapse, giving rise to distinct statistical properties which are used in star formation models, and by providing high density, filamentary structures where star-forming cores are preferentially located \citep{Scalo1998,Ferriere2001,MacLow2004,Kainulainen2009,Arzoumanian2011,Federrath2012,Schneider2012,Andre2014,Konstandin2016,Konyves2015,Federrath2016,Hacar2018,Mocz2018,Arzoumanian2019}. Understanding the structure, kinematics and the statistics (density and velocity dispersions, for example) of the MCs has therefore been of interest. The aim of this study is to extend upon our recent effort in \cite{Beattie2019}, herein called BFK19, to tie the fractal dimension, $\D$, to the physical properties of the MCs, e.g. to cloud length scales, and to the relations on observational data, e.g. between 2D cloud projections and 3D cloud data. 

In this study we present a new method for calculating the turbulent Mach number of the clouds, based purely upon two-dimensional (2D) projected cloud position-position (PP) data, i.e., the column density, $\Sigma$, which can be obtained by molecular lines, dust emission, or dust extinction observations. We also provide the first tests of the fractal dimension methods introduced by BFK19 using dust column density maps obtained from $Herschel$ flux maps of the Polaris Flare. First, we will discuss the importance of the turbulent Mach number, and the diverse role it plays in star formation.

\subsection{The Cloud Density and Turbulent Mach Number}\label{sec:intro_Mach}
The turbulent Mach number, $\M$, is a key ingredient for numerous star formation models \citep{Krumholz2005,Federrath2010,Hennebelle2011,Federrath2012,Federrath2013,Konstandin2016}. We make the distinction between the scale-dependent turbulent Mach number,
\begin{equation} \label{eq:Mach}
\Ml(\ell) = \sigma_v(\ell)/c_s, 
\end{equation}
where $\sigma_v(\ell)$ is the velocity dispersion of the cloud on length scale $\ell$, and $c_s$ is the sound speed, and the root-mean-squared (rms) Mach number, 
\begin{equation} \label{eq:rmsMach}
\M \approx \sigma_v(L)/c_s = \Ml(L),
\end{equation}
where $L$ is the cloud diameter, which corresponds to the outer scale of turbulence in our study \citep{Federrath2013}. For an isothermal cloud with purely turbulent dynamics, $\M$ sets the width of the log-normal cloud density distribution,
\begin{equation}
\sigma^2_s = \ln \left(1+b^2 \M^2 \right), 
\end{equation}
where the $s$ subscript denotes the variance of the normalised cloud density, $s = \ln(\rho/\rho_0)$, where $\rho_0$ is the mean density of the cloud and $b$ is the turbulent forcing parameter \citep{Padoan1997,Passot1998,Kritsuk2007,Federrath2008,Federrath2010,Konstandin2012b}. The dispersion has been studied extensively and there have been many modifications to account for 2D projections of the 3D cloud \citep{Burkhart2012}, thermal and magnetic pressures \citep{Padoan1997,Passot1998,Federrath2008,Price2011,Molina2012,Gazol2013}, and non-isothermal \citep{Nolan2015} and polytropic gases \citep{Passot1998,Li2002,Federrath2015}. Calculating the density dispersion is important for star formation models that predict the star formation rate (SFR) directly from integrating the density and free-fall time weighted cloud density distribution to determine the mass fraction of the cloud that could collapse into new stars \citep{Krumholz2005,Padoan2011,Hennebelle2011,Federrath2012,Kainulainen2014}. Beyond the density distribution, the turbulent Mach number may also play an important role in the distribution of cloud filament widths. 


\subsection{Filaments and the Turbulent Mach Number}\label{sec:intro_fil}
Filament structures have been observed in star-forming and quiescent clouds, and may play an important role in star formation, since star clusters and star-forming cores have been found to be preferentially located in them \citep{Andre2010,Menschikov2010,Schneider2012,Andre2014,Padoan2014,Arzoumanian2015,Federrath2016}. Interstellar filament widths are distributed with a peak at $\sim 0.1$pc, which seems to be a universal feature of filaments and has been found in both observations and simulations of star-forming clouds \citep{Arzoumanian2011,Juvela2012,Palmeirim2013,Andre2014,Smith2014,Benedettini2015,Kirk2015,Federrath2016,Federrath2016b,Smith2016,Arzoumanian2019}. The standard deviation of the filament width distribution is thought to be associated with the sonic scale, $\ell_s$, in the clouds \citep{Federrath2018}. The sonic scale marks the transition between supersonic and subsonic velocity dispersions, and is theorised to be at length scale, 
\begin{equation}
\ell_s = L \left[\frac{1}{\M} (1 + \beta)^{1/2} \right]^{1/p},
\end{equation} 
in the cloud, where $p \approx 1/2$ has been measured using both Galactic cloud observations and simulations, and $\beta = p_{\text{thermal}}/p_{\text{magnetic}}$ is the ratio between the thermal and magnetic pressures at the cloud diameter scale $L$  \citep{Larson1981,Solomon1987,Ossenkopf2002,Heyer2004,Kritsuk2007,Schmidt2009,Federrath2010,RomanDuval2011,Federrath2012,Federrath2016,Federrath2018}. Being able to measure the turbulent Mach number is thus essential for testing theories about the filament width distribution. The turbulent driving that the clouds undergo may lead to the formation of filaments through interacting planar shocks \citep{Federrath2016,Tokuda2018}. However these are not the only structures that are formed and the densities of turbulent clouds have been shown to respect a fractal geometry. 

\subsection{Fractal Cloud Structures}
Molecular clouds have a highly complex structure which includes sheets, filaments and dense cores. Observations of MCs through CO lines, dust emission as well as dust extinction show that they are organised into self-similar fractal structures, with substructures of clouds being continuously resolved, even at the highest spatial resolution achievable \citep{Falgarone1996,Stutzki1998,Chappell2001,Kauffmann2010,Schneider2013,Kainulainen2014,Rathborne2015}. There is a strong agreement between simulations and observations that the three-dimensional (3D) fractal dimension, i.e. the fractal dimension of the position-position-position (PPP) data of turbulent MCs falls between $\approx$ 2 and 3 (\citealt{Scalo1990,Elmegreen1996,Sanchez2005,Kowal2007,Federrath2009,Roman-Duval2010,Donovan-Meyer2013,Konstandin2016}; BFK19). However, where it falls between 2 and 3 depends upon the type of turbulent driving \citep{Federrath2009}, the rms $\M$ (\citealt{Konstandin2016}; BFK19), the length scales in the clouds, and the amount of shocks and filamentary structures in the cloud (BFK19). BFK19 also found that the fractal dimension is significantly higher in the column density map ($\approx 1.6$ in the high $\M$ limit) compared to 2D density slices ($\approx 1$ in the high $\M$ limit). In this study we show how by expanding upon the methods outlined in BFK19 one can utilise the fractal structure of the column density from the cloud, specifically the mass-length scaling, to measure $\M$, which is demonstrably an important quantity in star formation.\\

This study is organised into the following sections: In \S \ref{sec:sims} we discuss the six cloud simulations that we use to construct our new Mach number - fractal dimension $(\Ml - \D)$ relation. In \S \ref{sec:MLmethod} we summarise the fractal dimension method introduced by BFK19, including the key results. Next, in \S \ref{sec:MachNumberRelation} we derive the new relation, and discuss the applications and limitations. Then in \S \ref{sec:Application} we apply it to two quiescent subregions of the Polaris Flare to calculate the $\M$ based purely on the fractal geometry of the column density. We compare this with previous estimates of the $\M$ calculated in \cite{Schneider2013}. Finally, in \S \ref{sec:conclusion} we summarise our key findings.

\begin{figure*}
\centering
\includegraphics[width=\linewidth]{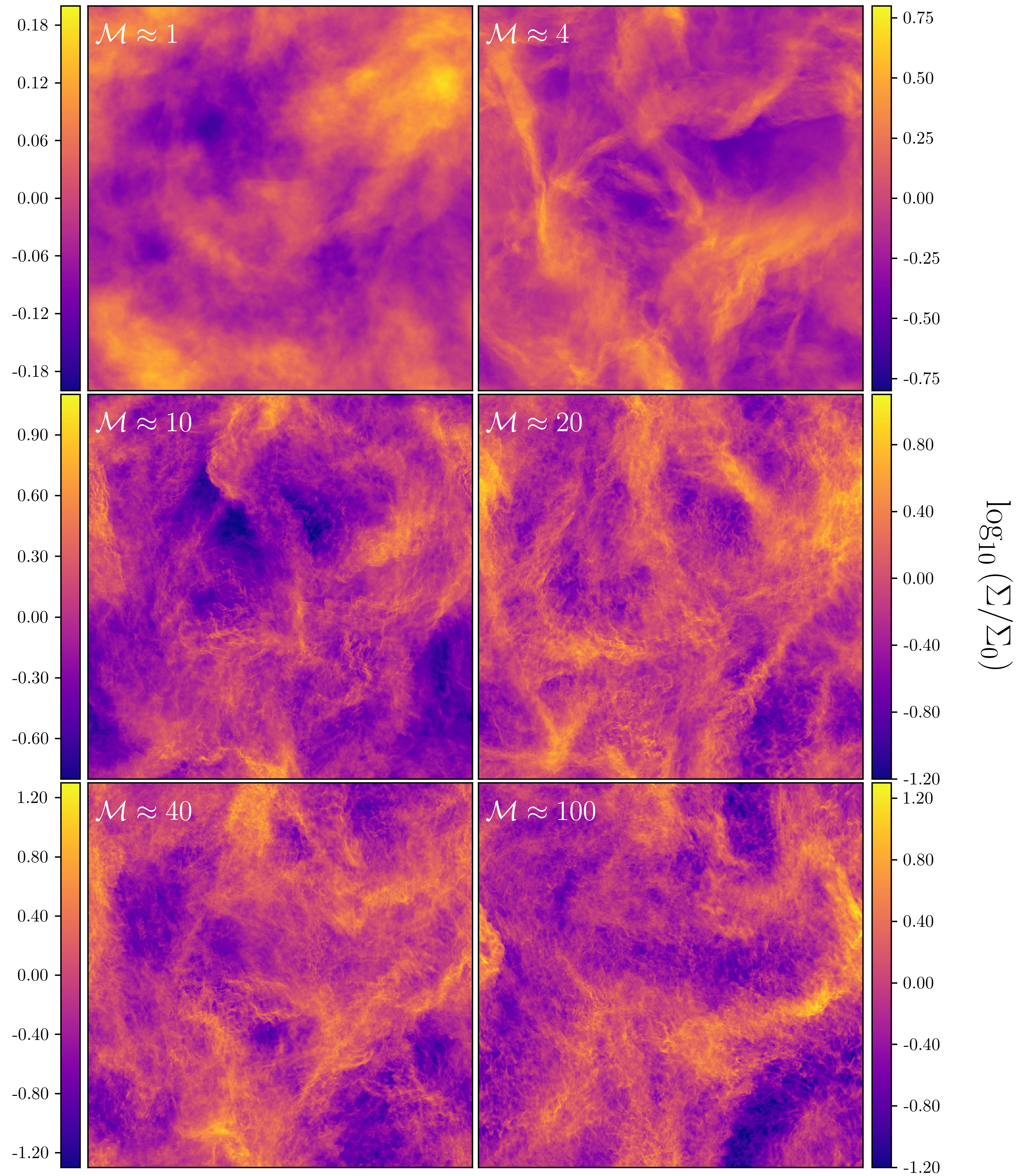}
\caption{The column densities, $\Sigma$, for the six simulations that we use to study the dependence of the turbulent Mach number, $\M$, on the fractal dimension, $\D$. Indicated in white is the root-mean-squared (rms) $\M$ of the simulation, ranging from the transonic $\M = 1$ clouds to the highly supersonic $\M = 100$ clouds. All column densities are shown at $t = 2\,T$, in the regime of fully-developed turbulence. The densities are shown in units of log average column density, $\Sigma_0$. An animation of the time evolution for the 2D projections (and 2D slices) is available in the online version of \protect\cite{Beattie2019}.}
\label{fig:Figure1}
\end{figure*}

\section{Turbulent Molecular Cloud Models} \label{sec:sims}
In this study we use six purely hydrodynamical simulations of quiescent (non-star-forming) molecular clouds, with no self-gravitation, to construct our $\M$ -- $\D$ relation. The parameter set of the molecular cloud models is listed in Table \ref{tb:simtab}. For each of the simulations we solve the compressible Euler equations,
\begin{align}
\frac{\partial \rho}{\partial t} + \nabla \cdot (\rho \mathbf{v}) &= 0, \\ 
\frac{\partial \mathbf{v}}{\partial t} + (\mathbf{v} \cdot \nabla)\mathbf{v} &= - \frac{1}{\rho} \nabla P + \mathbf{F},
\end{align}
where $\rho$ is the density, $\mathbf{v}$ the velocity, $P$ the pressure, following an isothermal equation of state, $P = c_s^2 \rho$, where $c_s$ is the speed of sound, and $\mathbf{F}$ is a Ornstein-Uhlenbeck (OU) forcing function that drives the turbulence through a mixture of solenoidal $(\nabla \cdot \textbf{F} = 0)$ and compressive $(\nabla \times \mathbf{F} = 0)$ modes. We choose a natural mixture of the two modes, $b \sim 0.4$. For further details we refer the reader to \cite{Federrath2010}, \cite{Federrath2018} and BFK19.

\begin{table}
\caption{Simulation parameters.}
\centering
\begin{tabular}{r@{}lccr@{}l}
\hline
\hline
\multicolumn{2}{c}{$\M$}& Native Simulation & Number of & \multicolumn{2}{c}{Time}\\
\multicolumn{2}{c}{$(\pm1\sigma)$} & Grid Resolution & Time Slices & \multicolumn{2}{c}{Interval} \\
\hline
1.01\, & $ \pm$ 0.05 & $1024^3$ & 71 & 2 $\leq t/T$&$\leq$ 9 \\
4.1\, & $ \pm$ 0.2 & $10048^3$ & 71 & 2 $\leq t/T$&$\leq$ 9  \\
10.2\, & $ \pm$ 0.5 & $1024^3$ & 71 & 2 $\leq t/T$&$\leq$ 9\\
20\, & $ \pm$ 1 & $1024^3$ & 71 & 2 $\leq t/T$&$\leq$ 9 \\
40\, & $ \pm$ 2 & $1024^3$ & 71 & 2 $\leq t/T$&$\leq$ 9 \\
100\, & $ \pm$ 5  & $1024^3$ & 71 & 2 $\leq t/T$&$\leq$ 9 \\
\hline
\hline
\end{tabular} \\
\begin{tablenotes}
\item{\textit{\textbf{Notes:}} Column (1): the rms turbulent Mach number of the simulation $\pm$ the 1$\sigma$ temporal fluctuations. Column (2): the native 3D grid resolution of each of the simulations. Column (3): the number of time slices that we use for temporal averaging. Column (4): the time interval in units of large-scale turnover times.}
\end{tablenotes}
\label{tb:simtab}
\end{table}

\begin{figure*}
\centering
\includegraphics[width=\linewidth]{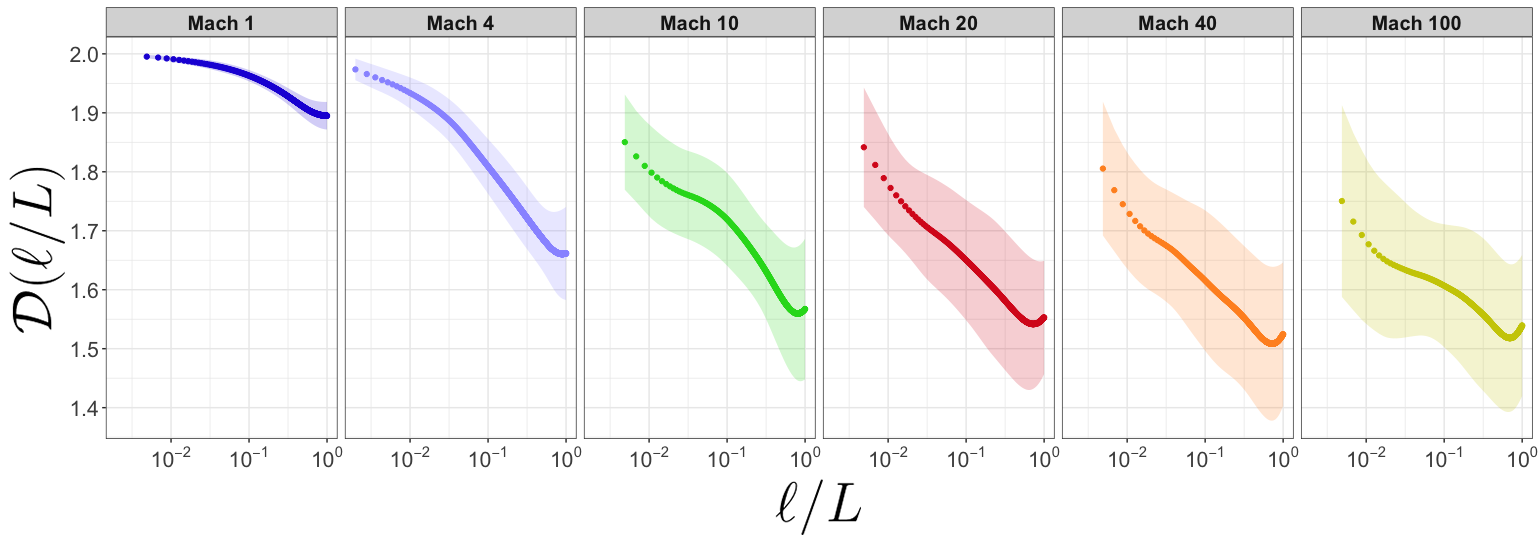}
\caption{Fractal dimension as a function of length scale, for each simulation. The rms Mach numbers for each simulation are indicated in the top panels. We see low $\M$ flows associated with high fractal dimensions and clouds driven at high $\M$ to low fractal dimensions. The plot suggests that there is a smooth transition between the high and low $\M$ limits, corresponding to a transition between space-filling and shock-dominated geometries in the cloud.}
\label{fig:FDcurves}
\end{figure*}

We run the six simulations with Mach numbers $\M=1,4,10,20,40$ and $100$, over seven turnover times $(7\,T)$ in the regime of fully-developed supersonic turbulence, established after $t \geq 2\,T$ \citep{Federrath2009,Price2010}. This gives us a wide set of $\M$ values to construct the $\Ml-\D$ relation, encompassing transonic, slightly compressible flows, all the way to highly supersonic, and highly compressible flows that are saturated with shocks \citep{Federrath2013}. We solve the Euler equations in a cube with periodic boundaries (for more details on the size of the grids we refer to BFK19) and use the same initial conditions for all simulations: a homogeneous medium at rest with $\rho(x,y,z,t=0) = \text{const.}$, $\mathbf{v}(x,y,z,t=0)=0$. The same random seed for the OU forcing function is used in all simulations, hence the only difference between them is the rms Mach number.

In this study we utilise the column density data, $\Sigma$, integrated along the $z$-axis. Figure \ref{fig:Figure1} shows a single snapshot in time, $t \approx 2 \, T$ of the  column density in each simulation. In the absence of magnetic fields our turbulence simulations are isotropic in a statistical sense, i.e. when averaged over time, e.g., \cite{Federrath2009,Federrath2010}. This lets us perform our study only on the $xy$ projections (column densities), whilst still being representative 2D projections through any viewing angle. 

\begin{figure*}
\centering
\includegraphics[width=\linewidth]{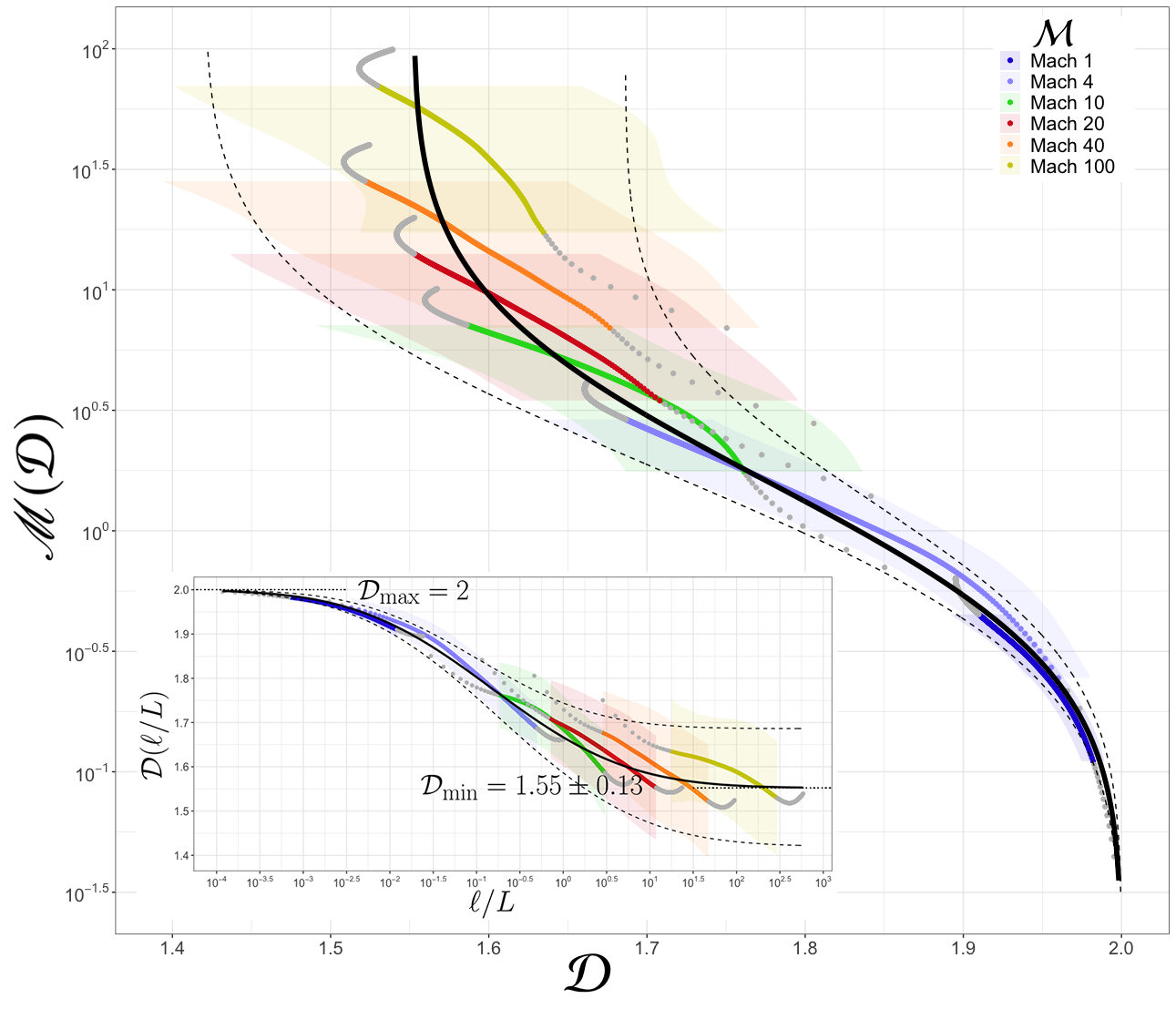}
\caption{ \textbf{Large panel:} The turbulent Mach number as a function of fractal dimension, fit on two decades of Mach number data ($\M = 1-100$). The different colours correspond to each of the six simulations, with rms Mach number indicated in the legend. The fit (Equation \ref{eq:M(D)}) is shown in black, with 1$\sigma$ uncertainties for the fit shown as dashed lines. \textbf{Small (inset) panel:} The fractal dimension as a function of length scale for the column density. This is a key result from \protect\cite{Beattie2019}, where it is discussed in detail. The black line is the fit shown in Equation \ref{eq:D(l)}. We use the same data, but transform the length scales into Mach numbers using a power-law scaling based on supersonic turbulence theory, $\ell \sim \Ml^2$, to create the $\Ml$ -- $\D$ relation \protect\citep{Burgers1948,Konstandin2012,Federrath2013}. }
\label{fig:MachFDRelation}
\end{figure*}

\section{Fractal Dimension Curves} \label{sec:MLmethod}
We follow the mass-length fractal dimension method outlined and discussed with detail in BFK19. This method allows us to calculate a mass-length $\D$ on each length scale in the cloud. It is important that we are able to access $\D$ on each $\ell$, since our aim is to relate $\D$ with $\Ml$ through $\ell$. We provide a summary of the method, and the key results below. Please note in BFK19 we use $\D_{\text{p}}$ to indicate the 2D projected (column density) fractal dimension but in this study we use $\D$ for simplicity.

\begin{table}
\caption{Fitting parameter values for $\D(\ell/L)$, shown in Equation \ref{eq:D(l)} and $\D(\Ml)$, shown in Equation \ref{eq:D(M)}.}
\centering
\begin{tabular}{cccccc}
\hline
\hline
$\Dmin \pm 1\sigma$ & $\Dmax$ & $\beta_0 \pm 1\sigma$ & $\beta_1 \pm 1\sigma$ & $\M \pm 1\sigma$ \\
\hline
$1.55 \pm 0.13$ & 2 & $0.46 \pm 0.08$ & $0.56 \pm 0.08$  & $4.1 \pm 0.2$ \\
\hline
\hline
\end{tabular} \\
\begin{tablenotes}
\item{\textit{\textbf{Notes:}} $\beta_0$, $\beta_1$ and $\Dmin$ are calculated in \cite{Beattie2019} using weighted non-linear regression. Column (1): $\Dmin$ is the minimum fractal dimension of the column density. Column (2): $\Dmax$ is assumed to be 2 for the maximum fractal dimension of the column density. This corresponds to completely space-filling flows on the 2D plane. Column (3): $\beta_0$ is a fitting parameter that corresponds to the translation of the complimentary error function over the $\ell/L$ axis. Column (4): $\beta_1$ is a fitting parameter that corresponds to the rate in which the complimentary error function changes between the high $\D$ and low $\D$ states. Column (5): The rms Mach number, $\M$ is measured by averaging over all rms Mach numbers from $2 \leq t/T \leq 9$ in the $\M=4$ simulation. The $1\sigma$ uncertainty is the standard deviation of the averaging process. We use $\M=4$ because the curves are in the Mach 4 simulation frame of reference, which is discussed in detail in \cite{Beattie2019}.}
\end{tablenotes}
\label{tb:parms1}
\end{table}

\subsection{Method Summary} \label{sec:methodsummary}
There exists a power-law scaling between the mass and length scales (size) in real MCs \citep{Larson1981,Myers1983,Falgarone1996,Roman-Duval2010,Donovan-Meyer2013}. To calculate the mass-length dimension we utilise this power-law scaling, where the mass, $m$ is given by
\begin{equation}\label{eq:ML}
m(\ell/L) \sim \left(\ell/L\right)^{\D},
\end{equation}
where we use the dimensionless $\ell/L$ for the length scales in the cloud, and $\D$ for the scaling exponent, i.e. the fractal dimension for the mass-length scaling relation. The constant of proportionality is $M$, the total mass, since when $\ell = L$, $m(1) \sim 1^{\D} = M$, in the unit system used here. We explore how the fractal dimension changes with spatial scale, i.e. we examine the $\ell/L$ dependence of the scaling exponent, $\D$,
\begin{equation}\label{eq:MLS}
m(\ell/L) = M (\ell/L)^{\mathcal{D}(\ell/L)}.
\end{equation}
We use Equation \ref{eq:MLS} to then define the fractal dimension at length scale $\ell_i/L$,
\begin{equation} \label{eq:MLS2}
    \mathcal{D}(\ell_i/L) = \frac{\log(m_i/M)}{\log(\ell_i/L)},
\end{equation}
where the $\ell_i/L$ length scale can be interpreted as all nested length scales up to the length scale $\ell_i/L$, i.e. $\ell_i / L = \left\{ \ell_0/L, \ell_1 /L , \hdots, \ell_i / L \right\}$, where $\ell_0/L < \ell_1/L < \hdots < \ell_i/L$ and $\ell_0/L$ is the smallest possible length scale in the cloud, and $m_i$ the corresponding mass at the $\ell_i / L$ scale, i.e., the total mass of the cloud on all scales less than and including $\ell_i / L$. This treats the cloud like a nested, hierarchical set of density objects, each with its own $\mathcal{D}$ and lets us probe how self-similar the cloud is over all length scales. If it is self-similar over a set of $\ell/L$, for example, the fractal dimension will not change as a function of $\ell/L$ over this region. This method also can be easily extended to explore the scale-dependent density structure of the cloud,
\begin{align}
    \frac{m}{(\ell/L)^2} &\sim (\ell/L)^{\D -2}, \\
    \Sigma &\sim (\ell/L)^{\D - 2},
\end{align}
to access the scaling in the column density, $\Sigma$.

To calculate the $\D$ described above we need to calculate the mass as a function of length scale. We do this by performing the following steps on the $\Sigma$ data from each simulation, following exactly the method outlined BFK19:

\begin{enumerate}
    \item Identify the coordinates of the maximum column density pixel $\Sigma_{\text{max}}$, \\[0.5em]
    \item expand a $(\ell/L)\times(\ell/L)$ square region centred on $\Sigma_{\text{max}}$, creating our length scale hierarchy, \\[0.5em]
    \item calculate the mass $m$, within each of the $(\ell/L)\times(\ell/L)$ squares, \\[0.5em]
    \item use the relation shown in Equation \ref{eq:MLS2} to determine the fractal dimension, $\D$, as a function of $\ell_i/L$, always using all length scales below $\ell_i/L$ to calculate $\D$ on $\ell_i/L$. \\[0.5em]
\end{enumerate}

We apply the four steps above on each of the 71 time slices in the interval $2 \leq t/T \leq 9$, the statistically fully-developed turbulence regime, averaging over them to construct a single curve for each simulation with 1$\sigma$ uncertainties.

\subsection{Key Results from the $\D(\ell)$ Curves}
In BFK19 we show that the fractal dimension curves from each simulation (shown in Figure \ref{fig:FDcurves}) can be combined together into the same common reference frame to create a composite fractal dimension curve. This lets us map the fractal dimension of $\Sigma$ over seven orders of magnitude in spatial scales, encompassing clouds undergoing subsonic to highly supersonic turbulent dynamics. After combining the curves we find that a complimentary error function is a good fit for $\D(\ell/L)$, which models a smooth transition between space-filling clouds ($\D=2$ in the 2D projection) and clouds saturated with planar shocks ($\D=1.55\pm 0.13$, which is a key result from BFK19). A simple power-law relation is not sufficient, because at both low and high rms Mach flows the fractal dimension curves begin to flatten out as they approach the high and low limits. The empirical fit is,
\begin{equation}\label{eq:D(l)}
\D(\ell/L) = \frac{\Dmax - \Dmin}{2}\erfc \Big( \beta_1\log \left( \ell/L \right) + \beta_0 \Big) + \Dmin,
\end{equation}
where $\erfc$ is the complimentary error function, $\Dmin$ and $\Dmax$ are the minimum and maximum fractal dimension, respectively, and $\beta_1$ and $\beta_0$ are fitting parameters, determined using nonlinear least squares and tabulated in Table \ref{tb:parms1}. This fit encodes the limits $\lim_{(\ell/L) \rightarrow  1 } \D(\ell/L) = \Dmin$ and $\lim_{(\ell/L) \rightarrow  0} \D(\ell/L) = \Dmax$ that we find in the data, and encompasses the smooth transition that we find between them. The composite curve data, along with the complimentary error function fit for $\D(\ell/L)$ are shown in the sub-panel of Figure \ref{fig:MachFDRelation}. The plot shows that the $\D$ of the column density is bounded between $\Dmax = 2$ and $\Dmin = 1.55 \pm 0.13$, where the former is assumed, and the latter is measured through the fitting process. Next, we use this curve to construct the $\Ml - \D$ relation.

\begin{figure*}
\centering
\includegraphics[width=\linewidth]{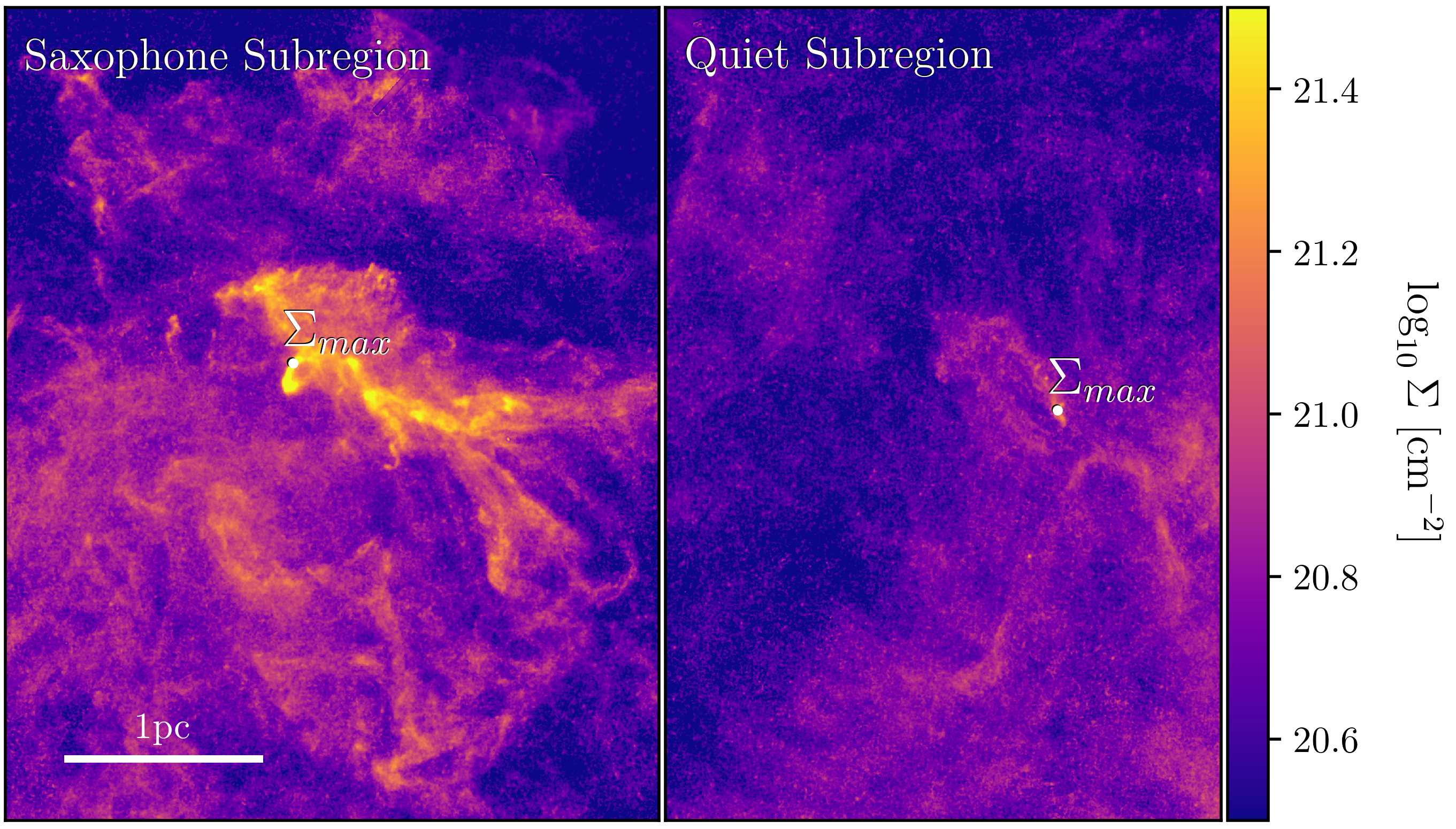}
\caption{The dust column density maps of the `saxophone' and `quiet' subregions of the Polaris Flare shown in $\log_{10}$ [cm$^{-22}$]. The maximum column density pixels, $\Sigma_\text{max}$ are indicated on the maps. These pixels were used to initialise the $\ell\times\ell$ expansion for the construction of $\D(\ell)$. The fractal dimensions of the regions are calculated to be $\D = 1.60 \pm 0.04$ and $\D = 1.76 \pm 0.05$, for `saxophone' and `quiet', respectively. This is consistent with \protect\cite{Beattie2019} where it is argued that low $\D$ values are associated with high $\M$ clouds, and high $\D$ values are associated with low $\M$ clouds.}
\label{fig:fig5}
\end{figure*}

\section{The $\Ml$ -- $\D$ Relation} \label{sec:MachNumberRelation}
After establishing the length scale dependence of the fractal dimension we may immediately ask how then does the fractal dimension change with the velocity dispersion of the cloud, since the velocity dispersion also depends upon length scale, as indicated in Equation \ref{eq:Mach}. This link is made available to us by approximating $\Ml(\ell)$ using scaling relations from models of supersonic turbulence. 

\subsection{Constructing the Relation}
Using the second-order structure function, $\SF_2(\ell/L) = \left\langle | \mathbf{v}(\mathbf{r}) - \mathbf{v}(\mathbf{r} + \ell/L)|^2 \right\rangle_{\mathbf{r}},$ where $\mathbf{v}(\mathbf{r})$ is the velocity of the cloud at position $\mathbf{r}$, and the operator $\left\langle \hdots \right\rangle_{\mathbf{r}}$ is an average over a large ensemble of spatial positions, one can construct the turbulent Mach number at length scale $\ell/L$ (using the definition of the second-order structure function and Equation 4.3 from \citealt{Konstandin2012}). This construction of the Mach number follows a power-law of the form,
\begin{equation}
\Ml(\ell/L) = \sqrt{\SF_2(\ell/L)/(2c_s^2)} \sim (\ell/L)^p,
\end{equation}
where $p \approx 1/2$ for supersonic turbulence and $p \approx 1/3$ for subsonic turbulence \citep{Kolmogorov1941,Burgers1948,Kritsuk2007,Schmidt2009,Konstandin2012,Federrath2013,Federrath2018}.  
We use the $p = 1/2$ case, $\ell/L \sim \Ml(\ell/L)^{2}$, to transform all length scales into Mach numbers. We set $\ell = L$ to find the constant of proportionality, $\M^{-2}$, i.e. on large scales in the cloud $\Ml$ is $\approx \M$ \citep{Federrath2013}. Hence the transformation is

\begin{equation} \label{eq:Mlmap}
\ell/L = [\Ml(\ell/L)/\M]^2.
\end{equation}

We apply this transformation to our $\ell/L$ composite data shown in the sub-panel of Figure \ref{fig:MachFDRelation}. This provides us with an estimate for $\mathcal{D}(\Ml)$,

\begin{equation}\label{eq:D(M)}
\D(\Ml) = \frac{\Dmax - \Dmin}{2}\erfc \Bigg[ 2 \beta_1 \log \left( \frac{\Ml}{\M} \right) + \beta_0 \Bigg] + \Dmin.
\end{equation}

We invert the equation to obtain $\Ml(\mathcal{D})$, which means that from measurements of the fractal dimension one can infer the scale-dependent turbulent Mach number,

\begin{equation}\label{eq:M(D)}
\log\Ml(\D) = \xi_1\left(\erfc^{-1}\left[ \frac{\D-\Dmin}{\Omega} \right] + \xi_2 \right),
\end{equation}
where 
$$\Omega = \frac{\Dmax-\Dmin}{2},$$
and
$$\xi_1 = (2\beta_1)^{-1} \hspace{2mm} \text{and} \hspace{2mm} \xi_2 = \xi_1^{-1}\log\M - \beta_0.$$
This corresponds to the turbulent Mach number on the length scale that $\D$ was measured, since $\Ml(\D(\ell))$. The values of the estimated and derived parameters for this fit are shown in Table \ref{tb:parms2}. 

\begin{table}
\centering
\caption{Derived parameter values for $\Ml(\D)$, shown in Equation \ref{eq:M(D)}.}
\begin{tabular}{cccc}
\hline
\hline
$\Dmin \pm 1\sigma$ & $\Omega \pm 1\sigma$ & $\xi_1 \pm 1\sigma$ & $\xi_2 \pm 1\sigma$ \\
\hline
$1.55 \pm 0.13$ & $0.22 \pm 0.07$ & $0.9 \pm 0.1$ & $0.2 \pm 0.2$ \\
\hline
\hline
\end{tabular} \\
\begin{tablenotes}
\item{\textit{\textbf{Notes:}} The values for $\Dmax$, $\beta_0$, $\beta_1$ and $\M$ can be found in Table \ref{tb:parms1}, which are used to derive the parameters in this table. Column (1): the same as column one from Table \ref{tb:parms1}, but included for completeness. Column (2): $\Omega = (\Dmax - \Dmin)/2$. Column (3): $\xi_1 = (2\beta_1)^{-1}$. Column (4): $\xi_2 =  \xi^{-1}_1 \log \M - \beta_0$.}
\end{tablenotes}
\label{tb:parms2}
\end{table}

\subsection{$\Ml$ -- $\D$ Relation Results}
In the main plot of Figure \ref{fig:MachFDRelation} we show $\Ml$ as a function of $\D$, derived from the column density data. The black line shows Equation \ref{eq:M(D)}, which is the equation we will use to convert fractal dimensions into Mach numbers. There are three main limitations to this method. The first is that the inverse complimentary error function has steep tails. This means that for low and high $\M$ the relation is extremely sensitive to small changes in $\D$. The second is that for high $\M$ the temporal fluctuations of $\D$ become significant, spanning over $> 0.2 \D$. This means that our relation will perform best at measuring regions of MCs with $10 \gtrsim \M \gtrsim 0.1$. Finally, since in our construction of the relation we use clouds driven by compressible, supersonic, isothermal and isotropic turbulence, with a natural mixture between solenoidal and compressive modes $(b \sim 0.4)$, the fit shown in Figure \ref{fig:MachFDRelation} may only work well, without modification, on quiescent clouds, without significant deviation from natural mixing, since $\D$ is sensitive to changes in driving \cite{Federrath2009}. Acknowledging these limitations, we now test the performance of the relation on $Herschel$ observations of subregions from the Polaris Flare cloud. 

\section{Application on quiescent subregions in the Polaris Flare Cloud} \label{sec:Application}
The Polaris Flare is a high Galactic latitude cloud which is located at a distance $\leq 150$pc \citep{Falgarone1998,Miville2010,Schlafly2014}. It has weak, but significant CO emissions in regions of higher hydrogen column density \citep{Falgarone1998,Meyerdierks1996,Miville2010}. There is no active star-formation in Polaris, only 5 starless cores were detected \citep{Andre2010,Thompson2010} that are most likely not gravitationally bound. Polaris is thus a perfect candidate for testing our new relation, which was calibrated upon quiescent cloud simulations. 

\subsection{Observational Data}
The Polaris region was observed as part of the $Herschel$ Gould Belt survey (HGBS, \citealt{Andre2010}) using the PACS and SPIRE instruments on-board $Herschel$. For all observational details, we refer to \citep{Andre2010,Menschikov2010,Thompson2010,Miville2010}. We employ publicly available level 3 data products produced with HIPE13 (Herschel Interactive Processing Environment) from the $Herschel$ archive. The angular resolution of the maps is 11.7$''$, 18.2$''$, 24.9$''$, and 36.3$''$ for 160 $\mu$m (PACS) and 250, 350, and 500 $\mu$m (SPIRE), respectively. For an absolute calibration of the maps (included in the SPIRE level 3 data), the Planck High Frequency Instrument (HFI) observations were used for the HIPE-internal zero-point correction task that calculates the absolute offsets, based on cross-calibration with HFI-545 and HFI-857 maps, including colour-correcting HFI to SPIRE wavebands, assuming a grey-body function with fixed beta. For the PACS 160 $\mu$m map, we obtained the zero-point correction following the procedure outlined in \cite{Bernard2010}.
Column density and temperature maps were then produced at an angular resolution of 18$''$, following the procedure outlined in \cite{Palmeirim2013} that employs a multi-scale decomposition of the imaging data and assumes a constant line-of-sight temperature. We performed a pixel-by-pixel SED (Spectral Energy Distribution) fit from 160 to 250$\mu$m, using a dust opacity law $\kappa_0 = 0.1 \times (\nu/1000GHz)^{\,\beta}$ cm$^2$ g$^{-1}$ with $\beta$ = 2 and assuming a gas-to-dust ratio of 100. This dust opacity law is commonly adopted in other HGBS publications and we refer to \cite{Andre2010} and \cite{Konyves2015} for further details. We estimate that the final uncertainties of the column density map are around 20--30 \%. \\
The resulting large, high-angular resolution (18$''$) hydrogen column density map traces structures between 0.01 to 8 pc \citep{Miville2010,Schneider2013}. For our study, we cut out the two subregions that have previously been used to investigate the link between the probability distribution function of column density (N-PDF) and $\M$ in Polaris \citep{Schneider2013}, the `saxophone' and `quiet' subregions, shown in Figure \ref{fig:fig5}. \cite{Schneider2013} made estimates of the turbulent Mach number for the two regions based on the CO\footnote{The CO data stem from $^{12}$CO 2$\to$1 and $^{13}$CO 1$\to$0 observations from the KOSMA and FCRAO telescopes  published in \cite{Bensch2003}.} velocity dispersion. The authors
calculated the $\M$ for each subregion using,
\begin{equation} \label{eq:S.Mach}
    \M = \left(\sqrt{3}\text{FWHM} \right) \Big/ \left(c_s \sqrt{8\ln 2} \right)
\end{equation}
where FWHM [km s$^{-1}$] is the full width at half maximum of the CO molecular line data, and under the LTE assumption the sound speed is $c_s \approx 0.188\sqrt{\text{T}_{\text{ex}}/10\text{K} }$, where $\text{T}_{\text{ex}}$ is the excitation temperature, $\text{T}_{\text{ex}} = 5.53\left[ \ln(5.33/\text{T}_{\text{CMB}}) + 1 \right]^{-1}$ and $\text{T}_{\text{CMB}} = 2.728$K is the temperature of the cosmic microwave background. The values for these estimates of $\M$ are shown in column (2) of Table \ref{tb:MachMeasurements}. 

\subsection{Application and considerations}
Using the method summarised in \S \ref{sec:methodsummary} we construct the fractal dimension curves for each of the two regions and then use Equation \ref{eq:M(D)} to calculate the Mach number\footnote{Our fractal dimension implementation on the Polaris Flare cloud is available here: \url{https://github.com/AstroJames/FractalGeometryofPolaris}}. The application differs in two ways compared to BFK19. (1) Here we use terminating boundary conditions, while on the simulation data we use periodic boundaries. Since the observational data is not periodic we terminate the $\ell\times\ell$ expansion on the boundaries. (2) We calculate the monofractal mass-length dimension instead of our length-dependent curves. Our method reduces exactly to the monofractal $\D$ when $\D(\ell = L)$, where $L$ is the largest scale of the $\ell\times\ell$ expansion. This means that we use all nested scales $\ell \leq L$ to calculate $\D$, as described in \S \ref{sec:methodsummary}. We do this because it allows us to calculate $\M$, the turbulent Mach number on the cloud scale, which is useful for constraining the 3D density PDF, calculating the star formation rate and estimating the sonic scale in the cloud (as discussed in \S \ref{sec:intro_Mach} and \ref{sec:intro_fil}, respectively). It also allows for comparison with previous estimates made for $\M$ in the `quiet' and `saxophone' subregions of the Polaris Flare molecular cloud using Equation \ref{eq:S.Mach}.

\subsection{$\M(\D)$ Results}
\begin{figure}
\centering
\includegraphics[width=\linewidth]{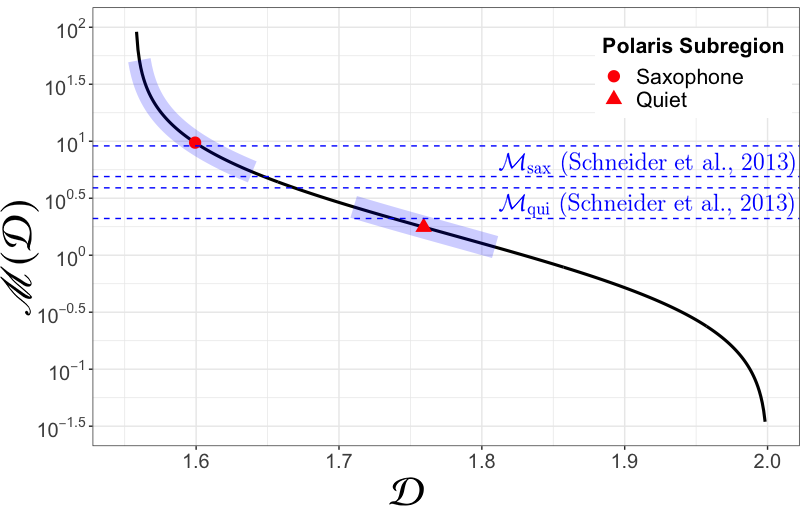}
\caption{The $\Ml$ -- $\D$ relation shown in black, with $\M$ measurements of the quiet and saxophone regions in the Polaris Flare cloud shown as red markers. The light-blue bands are the propagated $1\sigma$ uncertainties from the $\D$ measurements. The $\M_{\text{sax}} \sim 7$ and $\M_{\text{qui}} \sim 3$ regions, encompassed by the blue, dashed lines indicate the measured $\M$ using the CO velocity dispersion of the regions made by \protect\cite{Schneider2013}. We calculate $\M \sim 10$ and $\M \sim 2$ for the two regions, respectively, which is consistent with \protect\cite{Schneider2013} to within 1$\sigma$. The values of all $\M$ and $\D$ measurements can be found in Table \ref{tb:MachMeasurements}. }
\label{fig:fig4}
\end{figure}

In Figure \ref{fig:fig4} we show our $\M$ estimates for the two subregions of Polaris, indicated in red, and compare them with the $\M$ estimated by \cite{Schneider2013}, shown as blue-dashed regions. The $\M$ calculated by \cite{Schneider2013} and the $\M$ calculated in this study are shown in Table \ref{tb:MachMeasurements}. We find $\M = 2\pm 1$ and $\M = 10^{+55}_{-4}$, whereas \cite{Schneider2013} finds $\M 3 \pm 1$ and $\M = 7 \pm 3$, for the `quiet' and `saxophone' subregions of Polaris, respectively. These estimates are consistent to within 1$\sigma$. However the Mach number measurements are very sensitive to small changes in $\D$, especially for $\M \gtrsim 10$, where the relation becomes extremely steep. This translates into large, and not necessarily symmetric uncertainties, as shown for the `saxophone' subregion, in Table \ref{tb:MachMeasurements}. To understand why there may be differences between the $\M(\D)$ and previous estimates based on the CO velocity dispersion we turn to the calculation of $\D$.

To estimate $\M$ we first calculate $\D$, which is shown in column (3) of Table \ref{tb:MachMeasurements}. We find `quiet', the lower $\M$ region, has a $\D = 1.76 \pm 0.05$, and `saxophone', the higher $\M$ region, $\D = 1.60 \pm 0.04$. This is consistent with BFK19, who argues that for higher $\M$ flows we should expect lower $\D$, corresponding to the introduction of compressive shocks into a diffuse cloud with increasing $\M$, and previous studies have calculated $\D = 1.4 - 1.8$ for column densities \citep{Elmegreen1996,Elmegreen2004,Sanchez2005,Rathborne2015}. This suggests that the $\D$ values we calculate are reasonable, however, the `saxophone' region has a clear filamentary structure (see the high-density filament feature on the left column density map in Figure \ref{fig:fig5}), which has a density-length scaling relation $\rho \sim \ell^{-2}$, or $\Sigma \sim \ell^{-1}$, and will act to reduce $\D$ in the vicinity of $\Sigma_{\text{max}}$ \citep{Schneider2013,Federrath2016,Andre2017}. This may account for the slightly higher $\M$ that we estimate for `saxophone', compared to the Mach estimate in \cite{Schneider2013}. For the `quiet' subregion we slightly underestimate $\M$. This is because the column density (see the right column density map in Figure \ref{fig:fig5}) is diffuse, and lacks the shock structures that we see introduced between the $\M = 1$ and $\M = 4$ simulations in Figure \ref{fig:Figure1}. The different column density geometry in the `quiet' cloud may be due to a deviation away from the natural mixing of driving modes, towards stronger, compressive driving which we do not currently include in our relation, and which can change the fractal dimension up to $\sim 0.1\D$ for the mass-length method \citep{Federrath2009}.

\begin{table}
\caption{Mach number values for both Polaris Flare subregions.}
\centering
\begin{tabular}{cccc}
\hline
\hline
Subregion  & $\M$ (based on Eq. \ref{eq:S.Mach}) & $\D \pm 1\sigma$  & $\M(\D) \pm 1\sigma$  \\
\hline
Saxophone       & $ 7 \pm 3$  & $1.60 \pm 0.04$   & $10^{+55}_{-4}$ \\
Quiet           & $3 \pm 1$  & $1.76 \pm 0.05$   & $2 \pm 1$ \\
\hline
\hline
\end{tabular} \\
\begin{tablenotes}
\item{\textit{\textbf{Notes:}} Column (1): The subregion in the Polaris Flare cloud. Column (2): the estimated $\M$ from \cite{Schneider2013}, which is stated to have an error $\sim$30-40\%. Column (3): $\D$ calculated from our mass-length method. We take the $\D$ on the largest scale calculated, which reduces to the regular monofractal mass-length dimension. Column (4): The Mach number calculated from Equation \ref{eq:M(D)}, with 1$\sigma$ uncertainties propagated from column (3).}
\end{tablenotes}
\label{tb:MachMeasurements}
\end{table}

\section{Summary and Key Findings}\label{sec:conclusion}
In this study we construct a new empirical relation for the scale-dependent three-dimensional (3D) Mach number, $\Ml$, and the fractal dimension, $\D$, of the column density for turbulent clouds. We use the mass-length fractal dimension method introduced in \cite{Beattie2019} (BFK19) on six hydrodynamical cloud simulations, with root-mean-squared (rms) Mach number, $\M$, varying from $\M = 1$ to $100$. We apply the method on the column densities, with an example of the densities shown Figure \ref{fig:Figure1}, to construct $\D$ as a function of length scale, $\ell$. We then transform the cloud length scales to $\Ml$ using the scaling relation $\ell \sim \Ml^2$, for supersonic turbulence \citep{Burgers1948,Federrath2013}. Using this data we are able to construct $\D(\Ml)$, and finally $\Ml(\D)$. Using $\Ml(\D)$ we construct $\M$, where $\M \sim \Ml(L)$, and $L$ is the cloud scale, for the dust column density maps of two quiescent subregions from the Polaris Flare, that are termed `quiet' and `saxophone', studied earlier in \cite{Schneider2013}. We summarise our key findings below:

\begin{itemize}
\item We propose a new empirical relation for the scale-dependent Mach number and the fractal dimension of the column density, $$\log\Ml(\D) = \xi_1\left(\erfc^{-1}\left[ \frac{\D-\Dmin}{\Omega} \right] + \xi_2 \right),$$ as defined in Equation \ref{eq:M(D)} and plotted in Figure \ref{fig:MachFDRelation}, where $\xi_1 = 0.9 \pm 0.1$, $\xi_2 = 0.2 \pm 0.2$, $\Omega = 0.22 \pm 0.07$ and $\Dmin$, the fractal dimension of the column density in the high $\M$ limit is $1.55 \pm 0.13$. This relation allows for the calculation of $\M$ for clouds in the range $10 \gtrsim \M \gtrsim 0.1$. Very large and very low $\M$ are inappropriate for the model due to the steep tails in the inverse complimentary error function. \\[0.2em]
\item We use the mass-length fractal dimension method in BFK19 to calculate the $\D$ of the `saxophone' and `quiet' subregions of the Polaris Flare, shown in Figure \ref{fig:fig5}. We find $\D = 1.60 \pm 0.04$ and $\D = 1.76 \pm 0.05$ for `saxophone' and `quiet', respectively, consistent with the thesis that higher Mach number flows reduce $\D$, by turning diffuse, space-filling structures into compressive shocks and filaments. \\[0.2em]
\item Using $\D$ we estimate $\M$ for each of the subregions. We find `quiet' has a $\M \sim 2$ and `saxophone' has a $\M \sim 10$, shown in Figure \ref{fig:fig4}. This is comparable to the estimates made in \cite{Schneider2013}, $\M \sim 3$ and $\M \sim 7$ for the two respective subregions, but based on the CO velocity dispersion. The agreement between the Mach number estimate based on the CO velocity dispersion and based on our new fractal dimension relation, Equation \ref{eq:M(D)}, is acceptable, especially considering we do not account for how different types of turbulent driving influence the cloud geometries or how the presence of large filamentary structures, that locally scale the cloud by $\Sigma \sim \ell^{-1}$, act to reduce $\D$. \\[0.2em]
\item Our results suggest that the new empirical relation between the fractal dimension of column densities and 3D turbulent Mach number is a useful tool for extracting the Mach number purely from the structure and geometry of column density data from the cloud. 
\end{itemize}

\section*{Acknowledgements}\label{section:acknowledgments}
We thank the anonymous reviewer for the detailed and critical reading, which improved the study. J.~R.~B~acknowledges Tilly for the ever-present support. C.~F.~acknowledges funding provided by the Australian Research Council (Discovery Project DP170100603, and Future Fellowship FT180100495), and the Australia-Germany Joint Research Cooperation Scheme (UA-DAAD). R.~S.~K.~acknowledges support from the Deutsche Forschungsgemeinschaft via SFB 881, ``The Milky Way System" (sub-projects B1, B2 and B8). We further acknowledge high-performance computing resources provided by the Leibniz Rechenzentrum and the Gauss Centre for Supercomputing (grants~pr32lo, pr48pi and GCS Large-scale project~10391), the Partnership for Advanced Computing in Europe (PRACE grant pr89mu), the Australian National Computational Infrastructure (grant~ek9), and the Pawsey Supercomputing Centre with funding from the Australian Government and the Government of Western Australia, in the framework of the National Computational Merit Allocation Scheme and the ANU Allocation Scheme. The simulation software FLASH was in part developed by the DOE-supported Flash Centre for Computational Science at the University of Chicago. N.~S.~acknowledges support by the French ANR and the German DFG through the project "GENESIS" (ANR-16-CE92-0035-01/DFG1591/2-1).

\bibliographystyle{mnras.bst}
\bibliography{new.bib}

\label{lastpage}

\end{document}